\documentclass[conference]{IEEEtran}

\ifCLASSINFOpdf

\usepackage[pdftex]{graphicx}
\usepackage{subcaption}

\fi

\begin{document}
%
\title{Adaptive Multi-bit SRAM Topology Based Analog PUF}

%
%
%

\author{\IEEEauthorblockN{Sudarshan Sharma\IEEEauthorrefmark{1},
Dhruv Thapar\IEEEauthorrefmark{2}, Nikhil Bhelave\IEEEauthorrefmark{2} and
Mrigank Sharad\IEEEauthorrefmark{1}}
\IEEEauthorblockA{Department of Electronics and Electrical Communication Engineering\IEEEauthorrefmark{1},
Department of Electrical Engineering \IEEEauthorrefmark{2}\\
Indian Institute of Technology  Kharagpur, India\\
Email:sudarshansharma04@iitkgp.ac.in,
dhruvthapar97@gmail.com,
nikhil.bhelave@gmail.com,
mriganksh@gmail.com}}

\maketitle

\renewcommand\IEEEkeywordsname{Keywords}
\begin{abstract}

Physically Unclonable Functions (PUFs) are lightweight cryptographic primitives for generating unique signatures from minuscule manufacturing variations. In this work, we present lightweight, area efficient and low power adaptive multi-bit SRAM topology based Current Mirror Array (CMA) analog PUF design for securing the sensor nodes, authentication and key generation. The proposed Strong PUF increases the complexity of the machine learning attacks thus making it difficult for the adversary. The design is based on scl180 library.



\end{abstract}


%
\IEEEpeerreviewmaketitle

\section{Introduction}
%
%
%
%
Due to the whopping increase in the use of IoT devices, security has become one of the prime concerns. Physically Unclonable Functions (PUF) can be used as a low power alternative to secure sensor nodes. PUFs are mathematical models or physical structures that map the challenge words to corresponding responses which are governed by the uncertainties in the variation at the device level. Silicon PUFs have emerged as potential hardware cryptographic tools due to their ability to generate hardware-unique ‘responses’ to a given digital test word or ‘challenge’, by exploiting manufacturing process variations in circuit components in the IC. These random intrinsic variations are nearly impossible to replicate and therefore, PUFs provide extremely reliable hardware authentication and key generation. The popular applications of PUFs include hardware security and authentication such as secure RFIDs, IP protection in FPGAs and cryptographic key generation.

 
\subsection{Related Work}
The silicon realization of the PUFs (SPUF) is based on the random variations in dies across a wafer, and from wafer to wafer due to the process, temperature and pressure variations during the various manufacturing steps. The first implementation of PUFs on silicon is introduced in \cite{Gassend:2002:SPR:586110.586132}, where the delay variations of CMOS logic components are used to produce unique responses. In the delay-based PUF, the analog delay difference between two structurally identical parallel paths are compared which arises due to the manufacturing variations.
The Ring Oscillator (RO) PUFs are based on digital loops, they are easy to implement and possess higher reliability. However, RO PUFs are slower and consume larger power when compared to arbiter PUF (delay based PUFs) and depends heavily on the number of RO present. 


The low power current based PUF structure \cite{5938005} uses an automatic cut-off mechanism to stop the current flow after response evaluation to minimize the power consumption. A sub-threshold design based PUF \cite{5599030}
exploits the sensitivity due to process variation in deep sub-micron technologies,
the design consist of n stage CMOS multiplexers and delay circuits followed by an arbiter. To mitigate the power supply noise, switching noises and environmental variation a differential amplifier topology based PUF was introduced in \cite{Choi:2011:ICD:1960723.1960728}. 



Integrated circuit Identification (ICID) \cite{839821} is based on addressable MOSFET array which drives a load to generate random repeatable voltages based on the threshold voltage mismatches. SRAM cell and memory block based PUF exploiting the intrinsic process variations in read/write reliability of cells in static memories is implemented in \cite{10.1007/978-3-642-23951-9_27}. \cite{7870303} shows a PUF cell based on 2-T amplifier working in sub threshold region.  In \cite{5227103}, PUF design based on power grid resistance variation has been investigated. The response depends on the voltage drop at distinct locations of the IC occurred due to the introduction of a variety of stimuli. The concept based on the amplification of random transistor mismatch through two complementary current mirrors and a modified design with an addition of sense amplifier is discussed in \cite{7397840}. T Saha et al. \cite{7829547} proposed aging resistant, lightweight and low-power analog PUF which exploits the susceptibility of Threshold Voltage (Vth) of MOSFETs to process variations.


Almost every discussed PUF yield a single bit response for a unique challenge word which makes it easier for machine learning algorithms to model the PUFs with fewer challenge response pairs. We present an adaptive multi-bit SRAM topology based low power and highly robust analog PUF. The rest of the paper is organized in the following manner. Section II presents a discussion on Machine learning attacks while Section III describes the architecture of the proposed PUF. Section IV discusses the reliability, and finally, Section V concludes the paper.


\begin{figure*}
\centering
\includegraphics[width=\textwidth, height= 10cm]{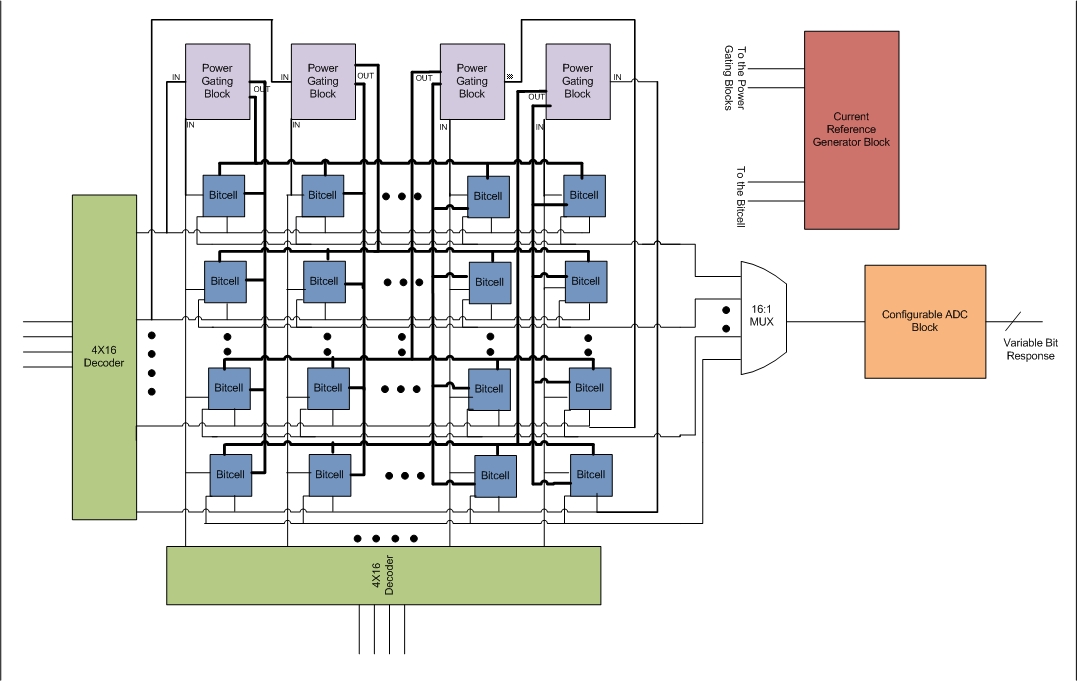}
\caption{\label{fig:Capture} Proposed PUF Architecture}
\end{figure*}


\section{Discussion on Machine Learning Based Attacks on PUF}
Most of the existing PUF circuits produce a single bit response for a challenge which can be modeled easily using machine learning algorithms. The proposed PUF adheres with the features of a Strong PUF following the property such as Many Challenges, Unpredictability and Unprotected Challenge Response Interface \cite{6800561}.
Ruhrmair et al. proposed various modelling attacks on PUFs in \cite{6800562}.
 The primary attack model consists of two steps, the first one requires finding a function with parameters which correctly describes the PUF’s challenge response behavior (or input-output behavior), followed by selection of a machine learning (ML) algorithm to train the parameters of the chosen function to improve its prediction quality using a large set of Challenge Response Pairs (CRPs) as training set. The best existing ML algorithms for attacks on PUFs involves Logistic Regression and Evolution Strategies. Other includes Support Vector Machines (SVM) and Neural Networks. The adaptive multi-bit response of the proposed PUF fails the SVM and logistic regression and makes it extremely difficult for heuristic-based methods like Evolution Strategies thus increasing the complexity of the attack.

\section{PUF Architecture}
Fig. 1 depicts the architecture of the proposed PUF. As illustrated in the figure, the two 4x16 decoders are used to convert the challenge word into bit-line and word-line which selects a particular bit cell from the entire array. Once the bit cell is selected, the power gating block routes the Current Mirror biasing to that specific bit cell. Its corresponding output through the MUX is processed by the configurable ADC block which produces the final variable multi-bit response of the challenge word.

\subsection{PUF Bit Cell}

The basic unit of the proposed SRAM topology based analog  PUF called bit cell comprises of two PMOS in cascode with two NMOS. All MOSFETs in the bit cell are minimum sized, 
hence increasing the probability of mismatch during manufacturing and also reducing the area required for the complete 16 x 16 bit-cell array. The output of the bit cell is chosen as the drain of PM2 and NM1 attributing to the symmetry. The cascode scheme used provides maximum amplification, and hence the mismatch on the extremities in even tenth of millivolts can be amplified.

PM2 and NM1 are biased through two transmission gates which are controlled via bit-line and word-line using the power gating blocks as seen in Fig 2.  The power gating block consists of two transmission gates(TG) controlled by the word line and the bit line, the TGs are tied with the current mirror biasing and the bit-cell as seen from Fig. 1. One power gating block is required per 16-bit cells (one column). This helps in switching off all unselected bit cells resulting in zero ideal state power consumption. Different switching schemes were tested and studied for various process corners. The channel length modulation results in a significant error while copying currents, especially for minimum size transistors. Thus a  current mirroring scheme becomes very important. Proposed configuration is shown in Fig. 4.

Due to variations in semiconductor manufacturing processes, corners are introduced in CMOS circuits. Five process corners dealing with process variations are typical-typical (TT), slow-slow (SS), fast-fast (FF), slow-fast (SF), fast-slow (FS). TT, FF and SS are not much of concern as both the MOSFETs get equally affected in one direction but other corners like SF, FS called as skewed corners may affect the response. Considering this, various current mirror configurations were tested and wide swing cascode current mirror worked well with the bit cell for all the corners with the perfect symmetrical output. The biasing current into the bit-cell using this particular current scheme varies around the center value (at no mismatch) of 4.3 $\mu$A.


Fig. 3 shows the response of various testbed used as current mirrors where X axis represents the referred variation modeled as the additional change in the threshold voltage of PM1. We observe that the ideal response is obtained through the proposed current mirror configuration. Moreover, the former has a higher gain and is perfectly symmetric compared to the reduced headroom cascode current mirror and simple cascode current mirror configuration.

\begin{figure}
\centering
\includegraphics[width=0.4 \textwidth]{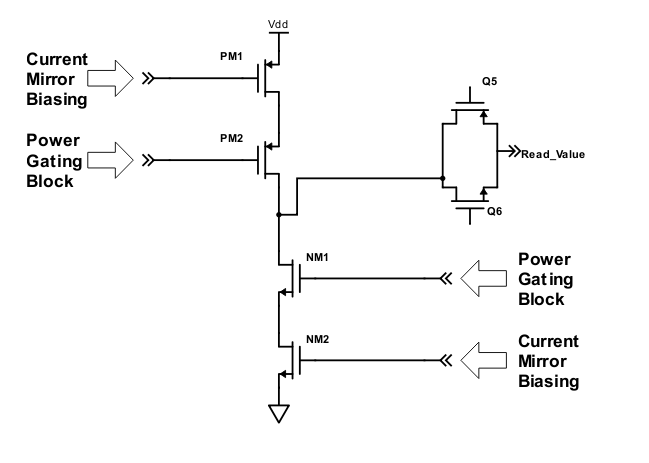}
\caption{\label{fig:}Bit Cell}
\end{figure}

\begin{figure}[ht!]
\centering
\includegraphics[width=0.35\textwidth]{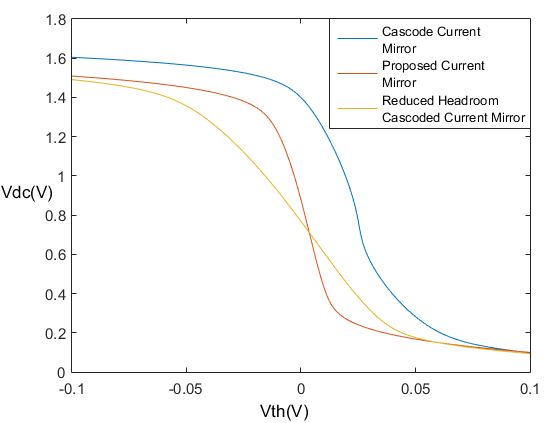}
\caption{\label{fig:All_schemes_1}Output Voltage Response for different Current Mirror Configurations}
\end{figure}

\begin{figure}[ht!]
\centering
\includegraphics[width=0.4\textwidth]{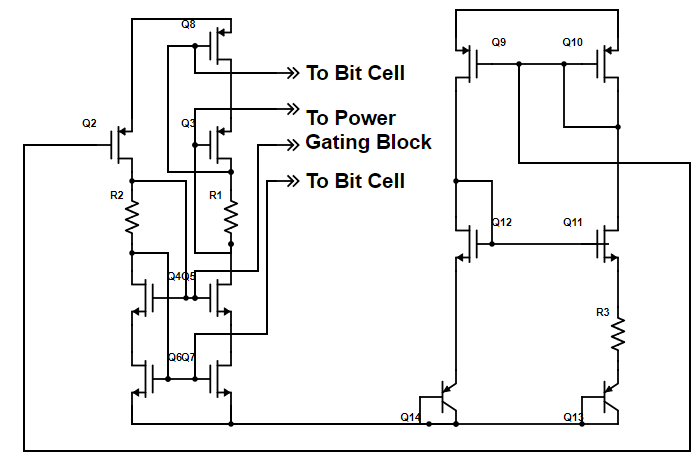}
\caption{\label{fig:} Wide Swing Cascode Current Mirror with PTAT Current Reference}
\end{figure}



\subsection{Switching Configurations}
The bit-cell zero static power in the idle state is achieved by turning the single bit cell ON only when it is selected and cutting OFF the supply rail for rest of the bit cell present in the entire array. Different switching configurations were tested, and their effects on the performance were studied in detail. The most viable option being the one in Fig. 5.
\begin{figure}[ht!]
\centering
\includegraphics[width=0.2\textwidth]{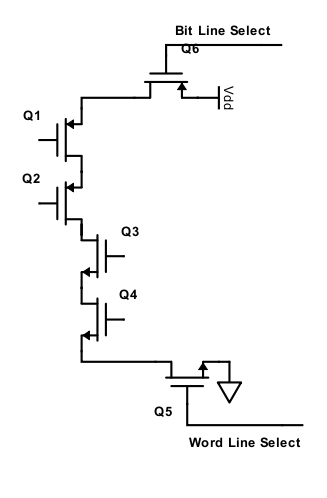}
\caption{\label{fig:up_down_switches}Naive Switching Configuration}
\end{figure}
The configuration does not require any additional circuitry to serve the purpose of switching. However, the response was hampered considering all process corners as a substantial amount of shift was observed in the output due to small voltage drops across the switching transistors.
To mitigate these effects control switches were removed from the bit cell and external control circuitry is employed (power gating blocks) which even drastically reduced the number of transistors per bit cell and works well with different process corners. The switching of PM1 and NM2 of the bit-cell will slow down the charging/discharging process from the pre-charged value to the desired value due to increased parasitic capacitance per bit line. Therefore, the two central MOSFETs, PM2 and NM1 are held with VDD and GND respectively which turns OFF the bit cell through the power gating block. Fig. 6 shows the comparison of the output voltage for different process corners in case of naive switching configuration and used switching configuration. 



\begin{figure}[ht!]
\centering
        \begin{subfigure}[b]{0.3\textwidth}
                \includegraphics[width=\linewidth]{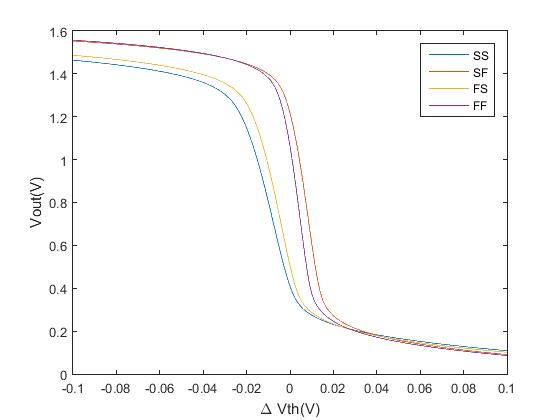}
                \caption{}
                \label{fig:gull}
        \end{subfigure}%

        \begin{subfigure}[b]{0.3\textwidth}
                \includegraphics[width=\linewidth]{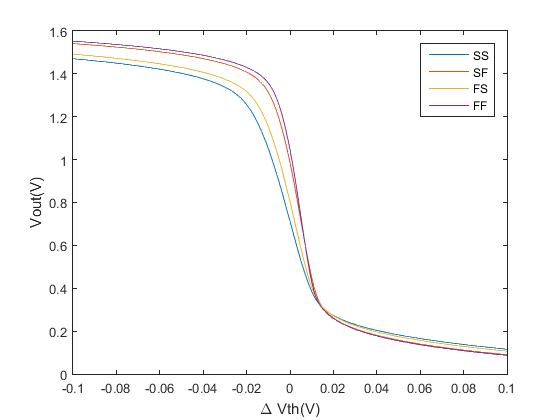}
                \caption{}
                \label{fig:gull2}
        \end{subfigure}%
        \caption{\label{fig1:}Output Voltage Variation for different Process Corners in (a) Naive Switching Configuration (b) Used Switching Configuration}
\end{figure}



\subsection{ADC scheme}
Due to the cascode configuration, the gain from the MOSFETs at the extremities to the output is very high thus it is highly probable that the response would hit either of the power rails, i.e., VDD or GND. The impact of the random process variations on circuit behavior can be studied through Monte Carlo(MC) simulation. The MC Simulation presented in Fig. 7 shows the output voltage distribution based on both the statistical variations, i.e., process and mismatch. As expected, the probability distribution is more skewed towards the power rails thus a varying multi-bit ADC scheme can exploit the utmost variations from the circuit. The final response is obtained through an adaptive multi-bit ADC scheme shown in Fig. 8. First, we define windows or regions as in the Fig. 8 using the Lloyd-Max Algorithm based on the probability distribution from the MC Simulation. The regions obtained are as follows: 

Region 1: [0,0.1451) 8 bit ADC

Region 2: [0.1451,0.6596) 7 bit ADC

Region 3: [0.6596,1.3308) 6 bit ADC

Region 4: [1.3308,1.6978) 7 bit ADC 

Region 5: [1.6978,1.8) 8 bit ADC

\begin{figure}[ht!]
\centering
\includegraphics[width=0.4\textwidth]{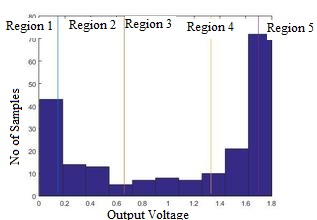}
\caption{\label{fig:}Monte Carlo Simulation considering Process Variation and Mismatch for Bitcell Output Voltage}
\end{figure}

The ADC selection logic unit governs bit precision of the conversion based on the regions mentioned above, and then the response is generated using the configurable single slope ADC scheme. V1, V2, V3, and V4 in Fig. 8 are the voltage values defining the windows.
\begin{figure}[ht!]
\centering
\includegraphics[width=0.4\textwidth]{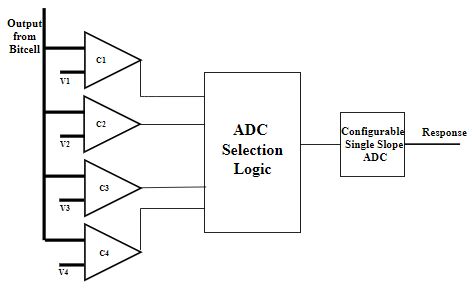}
\caption{\label{fig:}Configurable ADC scheme}
\end{figure}
\subsubsection{Configurable Single Slope ADC Scheme}
The single slope ADC technique has a series of advantages over the Flash and SAR ADC. The most important one being lower area and low power consumption. The single slope ADC technique designed in this case utilizes an external ramp circuit, a constant current source, three separate voltage comparators, a multiplexer, a free running timer, and a latching mechanism. The adaptive bit precision technique discussed is addressed using the variable frequency clock derived from the original clock along with variable counter values which are finally latched upon comparison during ADC operation. Fig. 9 describes the topology used to eliminate the Input Common Mode Range. The scheme is capable of correctly establishing ADC  operation for unknown analog voltage varying from 0 to VDD.
The two comparators one with NMOS as the input pair say Comparator A and the other one with PMOS as input pair say Comparator B includes offset cancellation scheme to prevent the results being skewed towards one direction. The decision of which comparator’s output to be used is performed by the third comparator i.e Comparator C based on the clause:
If the unknown voltage is higher than VDD/2, the output of Comparator A is selected else the output of Comparator B is used.
\begin{figure}[ht!]
\centering
\includegraphics[width=0.4\textwidth]{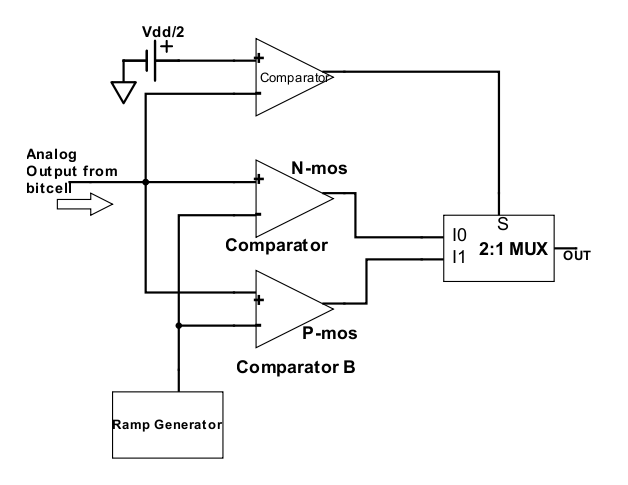}
\caption{\label{fig:}Single Slope ADC}
\end{figure}
\section{Results}
\subsection{Reliability with Temperature Variations}
The reliability of bit-cell output voltage is tested with temperature variations.  Fig 10 depicts the effect on the bit-cell output voltage as the temperature varies between $0\,^{\circ}\mathrm{C}$ and $60\,^{\circ}\mathrm{C}$. For simulation purposes, the referred variation is modeled as the change in the threshold voltage of the PM1 in the bit-cell.
\begin{figure}[ht!]
\centering
\includegraphics[width=0.35\textwidth]{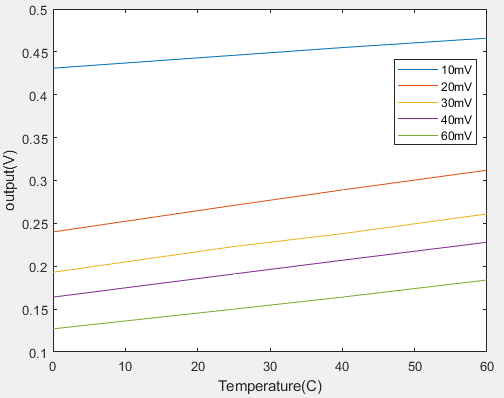}
\caption{\label{fig:}Bit-Cell output voltage variation across different temperatures at different variations in threshold voltages }
\end{figure}

\subsection{Comparison with other existing PUF schemes}
 PUFs operating in the sub-threshold region \cite{5599030} are shown to consume lesser power compared to those in super-threshold region but they exhibit higher delays thus reducing the speed of operation. The power gating block in the proposed PUF primarily contributes to the low power consumption without any hindrance in the operating speed. Table 1 provides the comparison with existing PUFs based on speed and power for 1 bit generation.
\begin{table}[ht!]
\renewcommand{\arraystretch}{1.3}
\caption{Power and Speed Comparison with PUF's for 1 bit generation}
\label{table_example}
\centering
\begin{tabular}{|c|c|c|}
\hline
PUF model & Power @ Speed of Operation & Energy / cycle\\
\hline
Super-threshold & 136.4$\mu$W @ 1GHz & 0.136 pJ\\
\hline
Sub-threshold & 0.047$\mu$W @ 1MHz & 0.047 pJ\\
\hline
ICID & 250$\mu$W @ 0.5MHz & 500 pJ\\
\hline
TV-PUF & 0.181$\mu$W @ 1 GHz & 0.0018 pJ\\
\hline
Proposed PUF & 306.54$\mu$W @ 6.4 GHz & 0.0478 pJ\\
\hline
\end{tabular}
\end{table}

\section{conclusion and future scope}
This paper discusses the architecture of an adaptive multi-bit Strong Analog PUF. Proposed Analog PUF is better than delay based PUFs considering unbiased responses. The output being independent of the actual layout design, the PUF is easy to fabricate due to its less circuit complexity. Furthermore, the PUF is shown to consume low power at a faster speed of operation.

The future work on this proposed PUF may include an evaluation based on uniqueness, uniformity and reliability after chip manufacturing. NIST tests can also be performed on the chip to compare it with more state of art PUF configurations.

\bibliographystyle{ieeetr}
\bibliography{citations}

\end{document}